\documentclass[12pt,manuscript]{aastex}

\shorttitle{SN 1994I}
\shortauthors{Clocchiatti et al.}

\begin{document}

\title{Late-Time {\em HST} Photometry of SN 1994I: Hints of Positron Annihilation Energy Deposition \altaffilmark{1}}

\author{
Alejandro Clocchiatti,\altaffilmark{2}
J. Craig Wheeler,\altaffilmark{3}
Robert P. Kirshner,\altaffilmark{4}
David Branch,\altaffilmark{5}
Peter Challis,\altaffilmark{4}
Roger A. Chevalier,\altaffilmark{6}
Alexei V. Filippenko,\altaffilmark{7}
Claes Fransson,\altaffilmark{8}
Peter Garnavich,\altaffilmark{9}
Bruno Leibundgut,\altaffilmark{10}
Nino Panagia,\altaffilmark{11}
Mark M. Phillips,\altaffilmark{12}
Nicholas B. Suntzeff,\altaffilmark{13}
Peter A. H\"oflich,\altaffilmark{14}
and
Jos\'e Gallardo\altaffilmark{15}
}

\altaffiltext{1}{Based in part on observations with the NASA/ESA \emph{Hubble
Space Telescope,} obtained at the Space Telescope Science Institute,
which is operated by the Association of Universities for Research in
Astronomy (AURA), Inc., under NASA contract NAS 5-26555. This
research is primarily associated with proposal GO--5777.}

\altaffiltext{2}{Pontificia Universidad Cat\'{o}lica de Chile,
Departamento de Astronom\'{\i}a y Astrof\'{\i}sica,
Casilla 306, Santiago 22, Chile; {aclocchi@astro.puc.cl}.}
\altaffiltext{3}{Department of Astronomy, The University of Texas at Austin,
Austin, TX, 78712; {wheel@astro.as.utexas.edu}.}
\altaffiltext{4}{Harvard-Smithsonian Center for Astrophysics, 60 Garden 
Street, Cambridge,
MA 02138; {kirshner@cfa.harvard.edu}, {pchallis@cfa.harvard.edu}.}
\altaffiltext{5}{Homer L. Dodge Department of Physics and Astronomy,
University of Oklahoma, 440 West Brooks, Room 100, Norman, OK
73019-2061;{branch@phyast.nhn.ou.edu}.}
\altaffiltext{6}{Department of Astronomy, University of Virginia,
P.O. Box 400325, Charlottesville, VA 22904-4325;{rac5x@virginia.edu}.}
\altaffiltext{7}{Department of Astronomy, University of California, Berkeley,
CA 94720-3411; {alex@astro.berkeley.edu}.}
\altaffiltext{8}{Department of Astronomy, Stockholm University, AlbaNova, SE-10691 Stockholm, Sweden; {fransson@astro.su.se}.}
\altaffiltext{9}{University of Notre Dame, Department of Physics, 225
Niewland Science Hall, Notre Dame, IN 46556-5670;
{pgarnavi@miranda.phys.nd.edu}}
\altaffiltext{10}{ESO, Karl-Schwarzschild-Strasse 2, Garching,
D-85748, Germany; {bleibund@eso.org}.}
\altaffiltext{11}{Space Telescope Science Institute, 3700 San Martin
Drive, Baltimore, MD 21218 (Also, Instituto Nazionale di Astrofisica
[INAF], Rome, Italy, and Supernova Ltd., Virgin Gorda, British Virgin
Islands).; {panagia@stsci.edu}.}
\altaffiltext{12}{Las Campanas Observatory, Casilla 601, La Serena,
Chile;{mmp@lco.cl}.}
\altaffiltext{13}{Texas A\&M University, Physics Department, College
Station, TX 77843; {suntzeff@physics.tamu.edu}.}
\altaffiltext{14}{
Department of Physics, Florida State University, 315 Keen Building
Tallahassee, FL 32306-4350; {pah@physics.fsu.edu}.}
\altaffiltext{15}{
Centre de Recherche Astronomique, \'Ecole Normale Superieure de Lyon,
France;{jose.gallardo@ens-lyon.fr}.}

\begin{abstract}

We present multicolor {\em Hubble Space Telescope (HST)} WFPC2 
broadband observations of the Type Ic SN 1994I
obtained $\sim$280~d after maximum light.
We measure the brightness of the SN and,
relying on the detailed spectroscopic database of SN~1994I, we 
transform the ground-based photometry obtained at
early times to the {\em HST} photometric system,
deriving light curves for the WFPC2
$F439W$, $F555W$, $F675W$, and $F814W$ passbands that 
extend from 7~d before to 280~d after maximum.
We use the multicolor photometry to build a
quasi-bolometric light curve of SN~1994I, and compare it
with similarly constructed light curves of other supernovae.
In doing so, we propose and test a scaling in energy and time 
that allows for a more meaningful
comparison of the exponential tails of different events.
Through comparison with models, we find that the late-time light 
curve of SN~1994I is consistent with that of spherically symmetric 
ejecta in homologous expansion, for which the ability to trap 
the $\gamma$-rays produced by the radioactive decay of 
$^{56}$\ion{Co}{0} diminishes roughly as the inverse of time squared.
We also find that by the time of the {\em HST} photometry, the 
light curve was significantly energized by the annihilation of 
positrons.

\end{abstract}


\section{Introduction}

Late-time light curves are an important piece of evidence for
understanding supernova (SN) explosions.
Conditioned by the always diminishing optical depth in the ejecta
a time is always reached when the energy from radioactive decay is radiated
on a time scale short compared with the dynamical time scale.
Radiative equilibrium is established and there is a simple 
connection between energy produced, trapped, and radiated.
The simplified physics permits a fairly direct connection between 
the light curve and the dynamical evolution of the expanding 
ejecta. This gives a more robust handle on some of the
basic physical parameters of the explosion.
How fast these late times are reached depend on the mass to energy ratio
of the ejecta.
For supernovae (SNe) without massive H envelopes they start a few weeks
after maximum light.

The main physical processes at play are the emission of 
$\gamma$-rays and positrons from radioactive decays in the
$^{56}$\ion{Ni}{0} $\rightarrow$ $^{56}$\ion{Co}{0} 
$\rightarrow$ $^{56}$\ion{Fe}{0}
chain \citep{cym69}, their interaction with the ejecta, and the
spectrum of the radiation produced by the thermalization 
processes and the radiative transfer in the expanding ejecta.
Some knowledge of the spectrum is important
because ideal bolometric light curves are seldom observed,
and astrophysicists generally must deal with observations that 
cover only a fraction of the radiated energy.
The decay of $^{56}$\ion{Ni}{0} into $^{56}$\ion{Co}{0} releases
energy in the form of  $\gamma$-rays, with an e-folding time of
8.8~d, while the decay of $^{56}$\ion{Co}{0} into $^{56}$\ion{Fe}{0} 
releases $\sim 96$\% of the energy in $\gamma$-rays and the
remaining $\sim 4$\% as positrons, with an e-folding time of 
111.3~d \citep{dandt98}.
Given a density structure and a chemical composition for
the expanding ejecta, the $\gamma$-ray transport problem is determined.
Positron interaction with the ejecta, on the other hand, strongly
depends on the presence, and geometry, of magnetic fields 
\citep{rlands98}.
These are, and will probably remain, essentially free parameters.
The escape of positrons from Type Ia supernovae (SNe~Ia) may be 
relevant to the diffuse 511~keV positron annihilation radiation 
from the Galactic bulge region \citep{kno05}.

Among SNe that explode through the gravitational collapse of 
the core, there are two distinct scenarios.
For SNe~II \citep[see][for a review of SN types]{filip97}, 
associated with explosions of massive stars that 
have retained most of their \ion{H}{0}/\ion{He}{0}
envelopes, and which are endowed with a relatively large 
mass-to-energy ratio, the ejecta are able first to trap the 
UV and optical-near IR photons for $\sim$100 days, and then to trap the
$\gamma$-rays from radioactive decays very efficiently for hundreds 
of days \citep[e.g.][]{cetal96a,eetal03}.
Since they trap essentially all of the energy from the radioactive 
decay of $^{56}$\ion{Co}{0} during the time when both 
$\gamma$-rays and positrons are being produced, their light 
curves do not provide an effective diagnostic for the balance 
of $\gamma$-ray versus positron energy deposition.
Something similar happens for many (though not necessarily
all) Type~Ib and Type~IIb explosions.

SNe of Type~Ic, on the other hand, associated with the explosion of 
stars that were massive while on the main sequence but that 
through individual and/or interactive binary evolution have lost 
all of the \ion{H}{0} and most of, if not all, the \ion{He}{0} 
of the outer layers, present more favorable conditions to study 
this problem.
Their ejecta have a mass, and/or mass-to-energy ratio, small enough 
to allow a major fraction of the $\gamma$-rays from $^{56}$\ion{Co}{0} 
to escape, potentially providing a cleaner window into the details 
of positron energy deposition.

The Type Ic SN 1994I in M51 presented a remarkable opportunity 
to build a useful late-time light curve for ejecta having both a 
low mass and a low mass-to-energy ratio.
It exploded in a nearby galaxy and thus reached a bright apparent 
magnitude at maximum.
Being in a host galaxy that is routinely observed by amateur 
astronomers meant, in addition, that the SN was spotted, 
reported, and spectroscopically confirmed well before maximum 
light \citep{petal94,setal94,cetal94}.
Consequently, it received extensive coverage both near maximum 
light and at later times.

An optical through near-infrared (IR) light curve extending $\sim$50~d 
after maximum light was prepared by \citet{schmidt94}, and a more 
extensive one, reaching $\sim$120~d after maximum, was presented 
by \citet{retal96}.
The SN was, in addition, the target of many other programs,
including radio and X-ray observations \citep{retal94a,ietal02}, 
and high-resolution spectroscopy of narrow absorption lines 
produced by significant amounts of foreground interstellar 
matter in M51 \citep{hetal95}.
Extensive spectroscopic data sets were collected and presented 
by \citet{fetal95} and \cite{cetal96b}. 
Finally, the light curve derived by \citet{schmidt94} was the subject 
of detailed theoretical studies to set constraints on the physical 
parameters of the progenitor of SN 1994I and its evolutionary path 
to explosion \citep{netal94,ietal94,wlyw95}.

In this paper, we present {\em Hubble Space Telescope (HST)} 
multi-wavelength broadband 
observations of SN~1994I, taken $\sim$280~d after maximum light
($\sim$300 days after explosive nucleosynthesis).
Combining these observations with those of \citet{retal96},
we build an optical through near-IR light curve that, via
detailed modeling, can be used to set constraints on the 
ejected mass and its dynamical and thermal evolution.
We introduce the new observations and photometry in 
\S~\ref{se:observations}. 
In \S~\ref{se:merging} 
we describe the problems that need to be solved in order
to combine the ground-based and space-borne observations, 
as well as the steps we took to construct our optical through
near-IR light curve, and in \S~\ref{se:lcurves} we present the 
results.
Our discussion in \S~\ref{se:discussion} gives some physical 
interpretation based both on simplified models of the ejecta 
and $\gamma$-ray radiative transfer, and presents a revision 
and extension of the detailed model CO21 \citep{ietal94}.
We summarize our conclusions in \S~\ref{se:conclusions}.

\section{Late-Time Observations and Photometry} \label{se:observations}

The late-time data on SN~1994I were obtained with {\em HST} under 
the Supernova INtensive Studies (SINS) General Observer Program.
On 1995 Jan. 15 (UT dates are used throughout this paper), about 280~d 
after maximum light, the WFPC2 on board {\em HST} was used to take 
five images of the SN in optical and near-IR passbands,
totaling 3200~s of exposure time.
The pointing of the telescope placed the SN on the Planetary 
Camera chip.
Table~\ref{ta:phot} gives additional details of the observations.
The images were reduced using the ``On-The-Fly Reprocessing'' pipeline,
provided by the Space Telescope Science Institute (STScI).

\begin{figure}
\epsscale{0.80}
\plotone{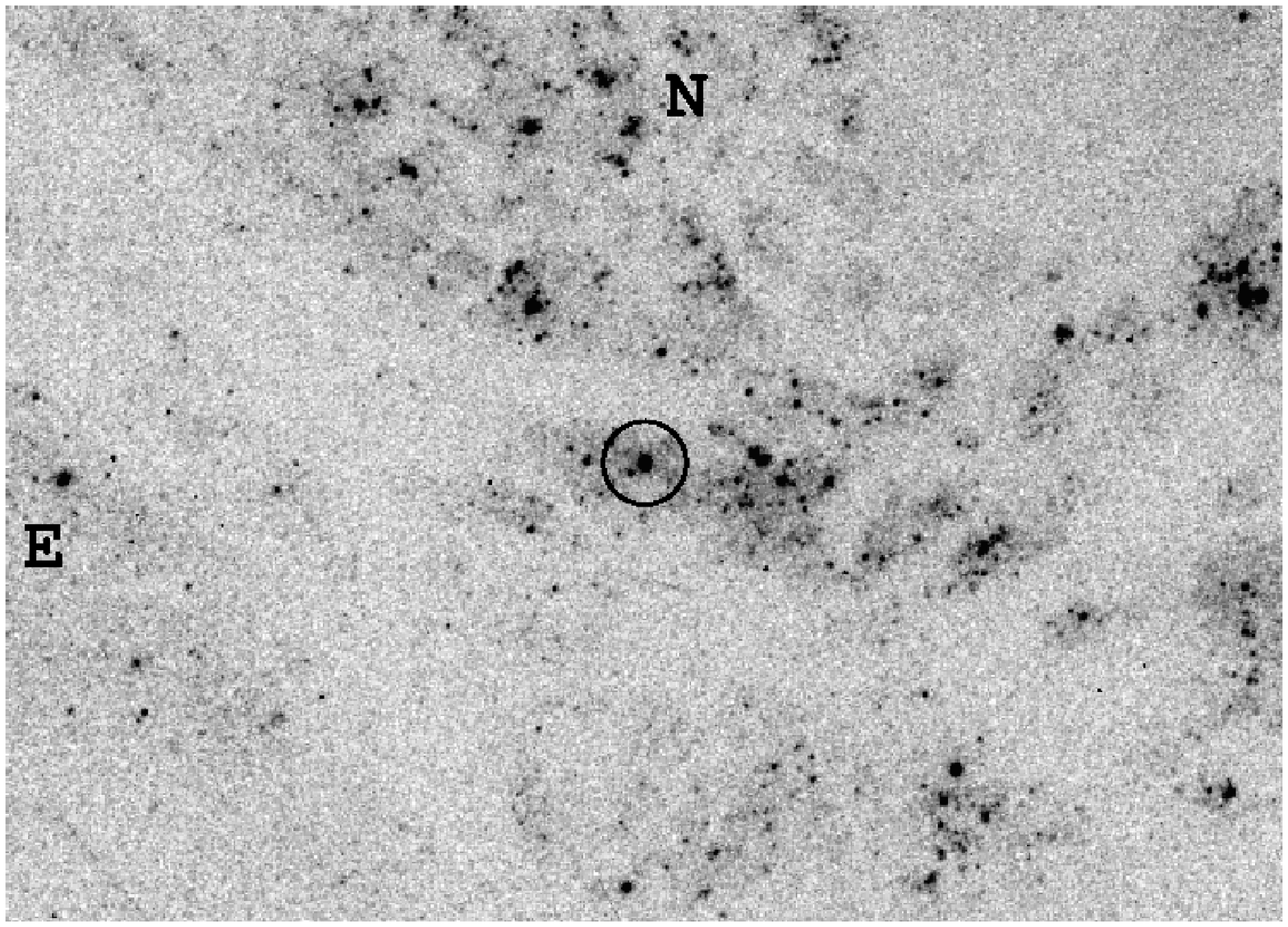}
\caption{\label{fi:finder}
{\em HST} image of SN~1994I and its surroundings.
The SN is at the center of the frame, inside the circle.
North is at the top and East is to the left. The width of the 
field displayed is about $28''$. The image shown is the 
combination of the three images taken on 1994 May 12, with 
the $F336W$ passband.
}
\end{figure}

\begin{figure}
\epsscale{0.80}
\plotone{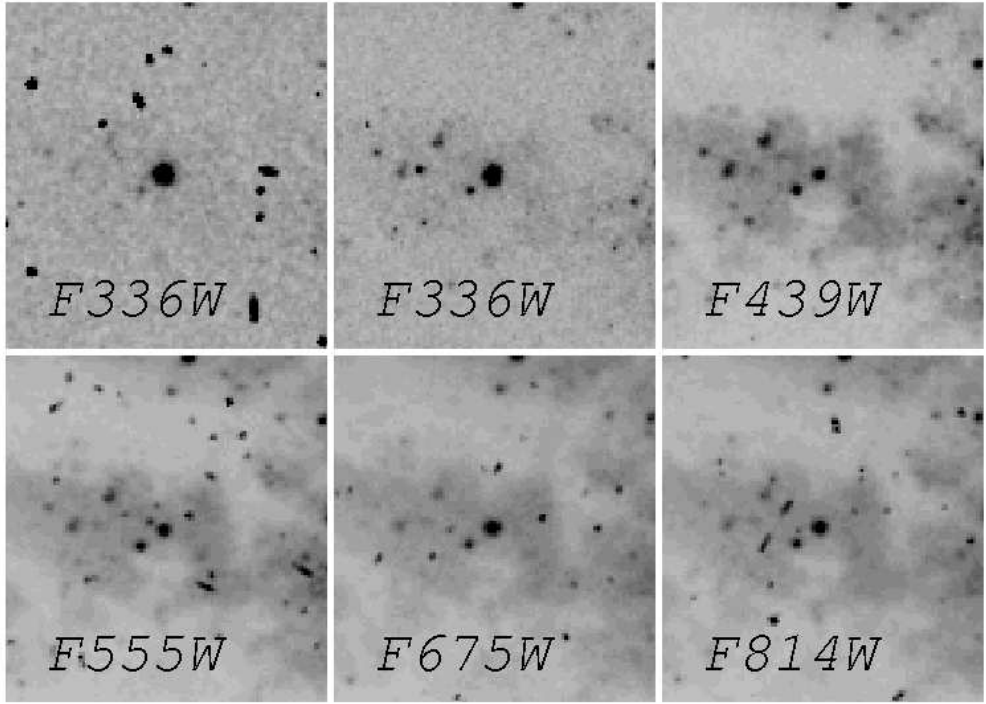}
\caption{\label{fi:stamps}
Stamps of the region around SN~1994I in the {\em HST} images.
North is at the top, and East is to the left. The width of each 
stamp is $4.6''$.
The passbands used are indicated by labels. SN~1994I is at the 
center of each stamp.
The top-left stamp corresponds to the $F336W$ exposure taken on 
1994 April 18, while the top-center stamp is an expanded view 
of the 1994 May 12 image shown in Figure~\ref{fi:finder}. The 
dimming of SN~1994I in the 24 elapsed days is clear.
The other images were taken on 1995 Jan. 15 
(see Table~\ref{ta:phot}).
}
\end{figure}

Careful inspection of the images revealed the SN at the expected 
position, amidst the complex structure of one of the inner arms of M51.
The right ascension and declination provided by the world coordinate 
system of the images was within $1''$ of the position found by 
high-resolution radio interferometry with the Very Large
Array \citep{retal94b}.
The identification was confirmed by analysis of two earlier 
sets of ultraviolet (UV) images also obtained with {\em HST}.
They were obtained with WFPC2, when the SN was close to maximum light,
using the $F336W$ passband.
One image, with an exposure time of 180~s, was taken on 1994 April 18, 
and a set of three images, with a combined exposure time of 1200~s,
was obtained on 1994 May 12.
The object identified as SN~1994I appears with very good signal 
in all images, and its flux diminishes by about a magnitude 
between the two epochs.
A finder chart showing SN~1994I in the {\em HST} UV images is given
in Figure~\ref{fi:finder}, and stamps of the region around the 
SN in the UV, optical, and near-IR images are given in 
Figure~\ref{fi:stamps}.

The background against which the SN appeared is complex, but the 
exquisite resolution of PC1 placed the SN more than 15 pixels 
away from the closest source.
The background light in the immediate vicinity of the SN appears 
to be relatively flat.
Point spread function (PSF) fitting photometry of the SN
was done using HSTPhot \citep{dol00},
which is calibrated with the zero points of \citet{hol95}
by normalizing the PSF magnitude to aperture photometry with 
a $0.5''$ radius.
The adopted version of HSTPhot had zero points and charge-transfer
efficiency (CTE) correction parameters updated in 2002 October.
The HSTPhot photometry of SN~1994I is given in Table~\ref{ta:phot}.

\section{Merging the Datasets and Estimating Extinction} \label{se:merging}

If the late-time photometry of SN~1994I presented in the previous 
section is combined with the ground-based photometry of 
\citet{retal96}, the result will be a multicolor light curve 
extending from $\sim 9$~d before to $\sim 280$~d after maximum light.
Very few SNe~Ic have a multicolor light curve extending so long.
Combining the datasets requires transforming the broadband 
observations between the different photometric systems used, 
and applying a correction for interstellar extinction.

The ground based observations of \citet{retal96} were taken with three
different combinations of instrument and telescopes but all of them
referred to the standard system of \citet{bessell90}.
We took the data from their Tables 4, 5, and 6, without making 
distinctions between datasets obtained with different telescopes,
and selecting the dates for which all the $B$, $V$, $R_C,$ and $I_C$ bands
had been observed.
When more than one datapoint existed for the same day we combined them
by a weighted average.
This approach means that systematic differences in the calibration of the SN
photometry due to the slightly different realizations of the standard system
are interpreted as an uncertainty in the observation.
The selected photometry is given in Table~\ref{ta:groundphot}.

\subsection{The Spectral Shape Corrections}

\label{sse:s_corrections}

Merging the space-borne and ground-based datasets calls for
careful consideration of the passbands used.
The differences in the SN flux measured by two different 
realizations of the same passband, evaluated in magnitudes, 
have been called ``spectral shape corrections,'' or $S$--corrections
\citep{setal02,ketal04,petal04}.
We generalize here the use of this name to the difference between the
flux measured by a ground-based passband and its space-borne 
counterpart.

Computing the $S$--corrections requires precise knowledge of 
the passbands' throughput and, ideally, spectra of SN~1994I at 
the specific dates of the photometry.
The spectroscopic evolution of SN~1994I was closely 
followed \citep{fetal95,cetal96b}, although the last spectrum 
available was taken only $\sim$147~d after maximum light,
more than 130~d before the {\em HST} photometry.
Regarding the passbands, those of the {\em HST} WFPC2 are well
characterized\footnote{See the URL ftp://ftp.stsci.edu/cdbs 
maintained by the STScI.}, but three slightly different 
realizations of the ground-based $BV(RI)_C$ system
were used by \citet{retal96}.
Hence, as a model for the ground-based system, we computed 
the $S$--corrections as if the standard system were that of 
\citet{bessell90}, the one to which \citet{retal96} attempted 
to transform.
We estimated an uncertainty for these $S$--corrections
by using different realizations of the Bessell passbands
at assorted optical/near-IR imagers at the Cerro Tololo 
Interamerican Observatory and the Kitt Peak National Observatory.
Note that we transformed the curves of \citet{bessell90} 
from energy-based to photon-based units, and, for the
other ground bassed passbands, we factored in a typical
CCD quantum efficiency, the spectral response of two 
aluminum-coated surfaces, and the typical response of the sky 
telluric absorption bands if needed.

With this database of passbands and spectra, we explored whether 
it was better to transform the photometry from the space-borne 
to the ground-based system, or vice-versa. 
We found that because spectra are available at the times of the 
ground-based observations, it was preferable to convert the 
ground-based photometry to the space-based system.
We computed $S$--corrections for the ground-based photometry
using the spectra of \citet{fetal95} and \citet{cetal96b},
and passbands as described above. We fit smooth polynomials 
to their time variation and used them to interpolate the 
$S$--correction and its uncertainty for the \citet{retal96} 
observations.
The results are included in Table~\ref{ta:groundphot}.

\subsection{Interstellar Extinction}

\label{sse:extinction}

All of the photometric and spectroscopic observations of 
SN~1994I indicate that it was heavily reddened by interstellar 
matter in M51.
Different approaches to estimating $A_V$ have included comparison 
with detailed theoretical models of light curves \citep{ietal94}
and spectra \citep{betal96,betal99,metal99}, and high-resolution 
spectroscopy of the narrow absorption lines produced by the gaseous 
phases of the interstellar matter \citep{hetal95}.
Values of $A_V$ from 0.9 up to 3.1 have been inferred, with a 
preferred range roughly between 1 and 2.
We follow \citet{retal96} in adopting $A_V = 1.4 \pm 0.5$ mag.

This value of $A_V$ must next be translated into extinction for 
the different broadband filters.
Typically an interstellar reddening law is adopted and scaled by
the value of $A_V$, and the monochromatic fluxes obtained from 
the photometry are then corrected, applying the extinction value 
that the adopted law prescribes for the effective wavelength
of the passbands.
This approach, however, may lead to systematic errors when 
applied to objects with strong and variable emission lines.
The effect is typically small, but it is systematic with time,
tending to change the overall shape, and/or slope, of a light curve.
Since the extinction affecting SN~1994I is large, the issue 
merits consideration.

Again, the series of spectra of \citet{fetal95} and 
\citet{cetal96b} allow us to address this issue. 
Using these spectra,
a typical interstellar extinction law \citep{ccm89}, with $R_V = 3.1$,
the transmittance of the passbands described earlier, and typical
instrumental sensitivities, it is possible to 
compute the extinction affecting the spectra and its 
variation with time.
The estimated extinction values and their
uncertainties are included in Table~\ref{ta:groundphot}.

The lack of a spectrum of SN~1994I some 280~d after maximum
affects the estimates of interstellar extinction at the time, but
the effect is small.
Most of the variations that strongly affect the $S$--corrections 
at late times (changes of the emission lines in the near-IR) have 
a minor impact on the extinction.

We computed the extinction for the passbands of the {\em HST} 
WFPC2, fit smooth polynomials to the late-time trends, and 
extrapolated those from the latest observed spectrum out to 
the latest photometric point.
To account for the uncertainty of the extrapolation, we increased 
the typical uncertainty of the fits by a factor of three for the 
last point.
The extinction computed for the {\em HST} observation
is included in Table~\ref{ta:phot}.

The uncertainties due to the unknown details of the 
ground-based standard system realization
and the lack of a very late-time spectrum
are much smaller than the uncertainty of the overall 
extinction adopted \citep[$\Delta A_V \approx 0.5$ mag according to][]{retal96}.
The changes in the light curves associated with the 
variation of the broadband extinction with time fall well 
within $\Delta A_V$ as well.
However, computing the extinction from the evolving spectra 
and the expected passbands is the correct procedure for 
nonstellar spectra, and provides light curves that, for a 
given velue of $A_V$, are free from the systematic effects
related to the time variation of the extinction in each passband.

Two additional sources of uncertainty should me mentioned. First, the
interstellar extinction law may not be the standard one.
What can be said in this case is that, if the interstellar extinction towards
SN~1994I follows the functional form of \citet{ccm89} with a different value of
$R_V$ the difference will be smaller than the uncertainty in $A_V$, even for the
more extreme values of $R_V$ that they consider.
Second, it is possible that at least a fraction of the extinction is produced in the
local environment of the supernova. Since this local environment will be
eventually affected by the evolution of the ejecta, a change of the foreground
extinction with time would be conceivable. With the data available, it is not
possible to set quantitative limits to this possibility.

\section{The Light Curves} \label{se:lcurves}

\begin{figure}
\epsscale{0.80}
\plotone{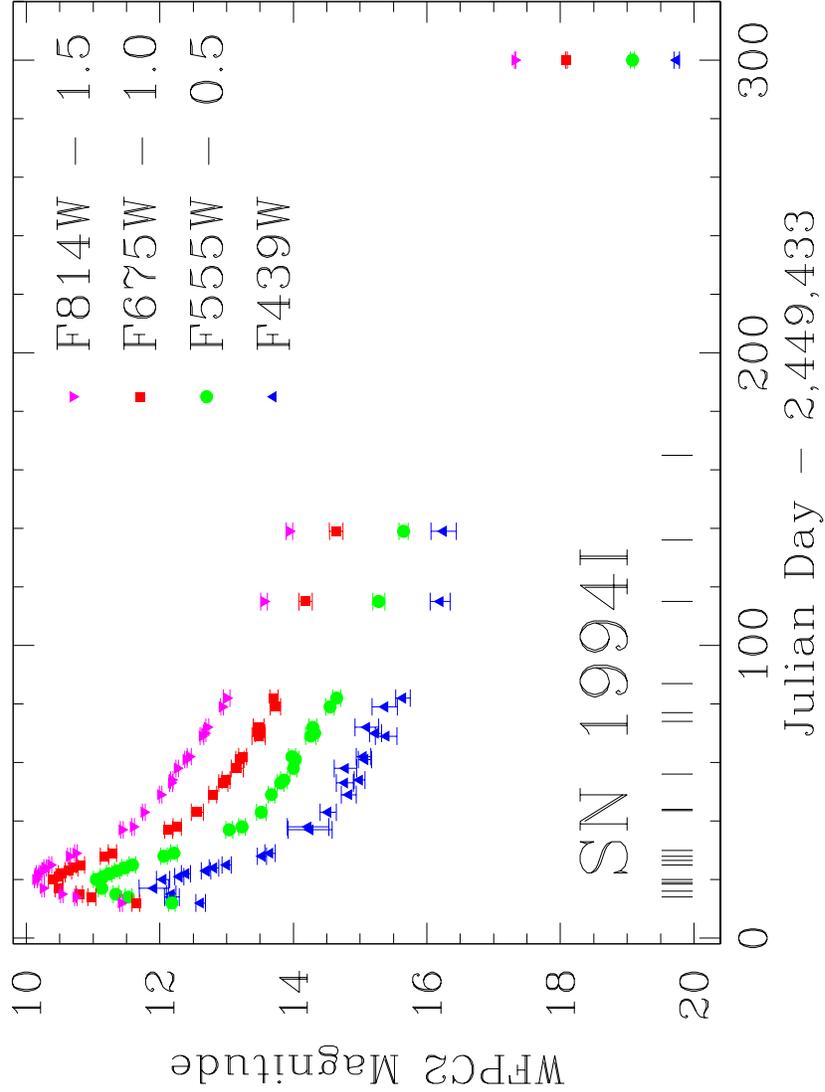}
\caption{\label{fi:phot}
{\em HST} WFPC2 broadband unreddened photometry of SN~1994I 
(the last point, at JD $-$ 2,449,433 $\approx$ 300),
and ground-based $B$, $V$, $R_C$, and $I_C$ photometry 
\citep{retal96} unreddened and translated to the WFPC2 
passbands.
The vertical marks above the lower horizontal axis indicate
the times at which the spectra of \citet{fetal95} and
\citet{cetal96b} were taken.
The shift in time corresponds to the explosion date ($JD = 
2,449,433$) estimated by \citet{ietal94}.
}
\end{figure}

\begin{figure}
\epsscale{0.80}
\plotone{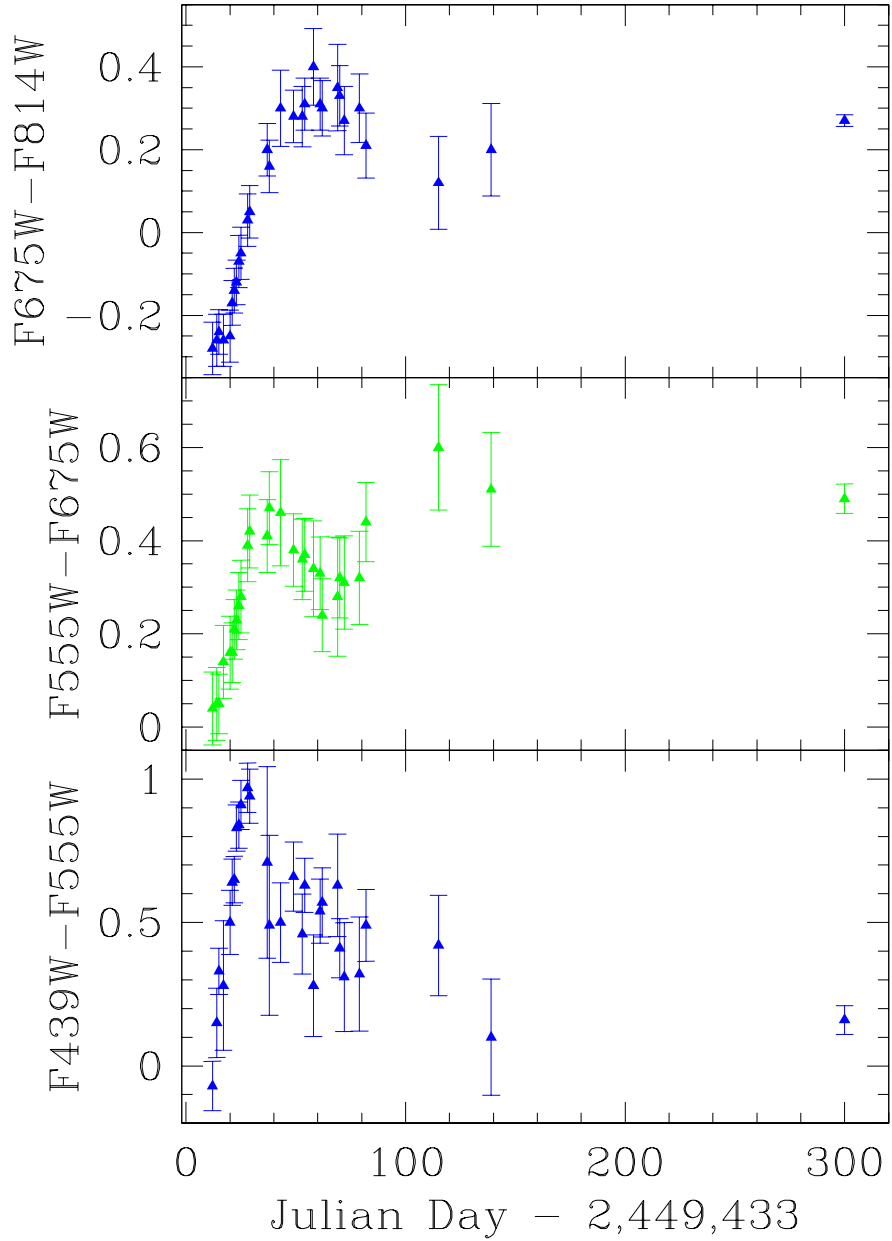}
\caption{\label{fi:colors}
{\em HST} WFPC2 broadband unreddened colors of SN~1994I 
in the HST passbands
computed from the photometry of Figure{\protect{\ref{fi:phot}}}.
The shift in time corresponds to the explosion date ($JD = 
2,449,433$) estimated by \citet{ietal94}.
}
\end{figure}

The resulting light curves of SN~1994I in the space-based system 
are shown in Figure~\ref{fi:phot}, while the evolution of the photometric
colors is given in Figure~\ref{fi:colors}.
The late-time point is consistent with the decrease in brightness and
a flattening of the light curve expected from models that fit the post maximum decline (see below), and as suggested by the ground-based data earlier than $\sim$80~d after maximum.
The $(RI)_C$ magnitudes of the last two ground-based observations
(at $\sim$115 and $\sim$140~d after maximum), and the $B$ and 
$V$ magnitude of the last ground-based observation,
are not consistent with this trend.
They appear to be slightly overluminous.
The observed colors indicate little or no change of the overall spectral shape since
the latest ground-based photometry.

With all the photometry transformed into the same system, it is 
possible to convert the observed single-bandpass light curves into
monochromatic fluxes at the effective wavelengths, integrate 
them over wavelength, and obtain a new version
of the ``$BV(RI)_C$'' light curve of SN~1994I\footnote{Even 
though it was calibrated using the space-borne WFPC2
passbands, we prefer to name the light curve using the more 
compact designation of their equivalent ground-based passbands.}.
We used the zero points for the WFPC2 passbands given by 
\citet{hol95}, and integrated the monochromatic fluxes using a
simple trapezoid rule.
We did not make any assumptions regarding the behavior of the flux 
beyond the limiting wavelengths of our coverage;
flux beyond the limits of the $B$ and $I_C$ passbands was ignored.
The distance of M51 was assumed to be 8.39 Mpc, based on the 
luminosity function of planetary nebulae \citep{fcj97}.

The $BV(RI)_C$ light curve of SN~1994I is shown in 
Figure~\ref{fi:BVRI_1}, together with similarly constructed 
light curves of the Type Ic SNe 1990B \citep{cetal01}, 1998bw 
\citep{petal01,soetal02,cetal07}, and 2002ap \citep{tetal06}.
The latter two have been modeled with ejecta of low 
mass-to-energy ratio \citep{ietal98,metal02},
but fairly large ejected mass, and have also sometimes been 
called ``hypernovae.''
As a benchmark of absolute brightness and evolution timescales,
the $BV(RI)_C$ light curve of the normal Type Ia SN~1992A (N. 
Suntzeff, 2006, private communication) is also given.

\section{Discussion} \label{se:discussion}

\subsection{Simple Comparisons} \label{sse:simple}

The light curves in Fig.~\ref{fi:BVRI_1} illustrate the 
wide range that SNe~Ic span in absolute brightness, and 
show that the diversity cuts across observational borders 
proposed to classify them, like the supernova or hypernova 
categories.
Fig.~\ref{fi:BVRI_1} also shows that SN~1994I
was fairly luminous reaching more than 50\% of
the brightness of normal Type Ia SN.
%
A foreground extinction a few tenths of a magnitude larger,
which is consistent with model light
curves, spectra, or high-resolution spectra of the narrow
\ion{Na}{0}~$D$ absorption lines, would
make it as bright as SN 1992A.
The $BV(RI)_C$ light curves also confirm that, similar to 
the behavior close to maximum light, the time scale for 
brightness variation of SN~1994I at late times was fast.
The early exponential tail of SN~1994I is faster than that of
the other Type~Ic SNe, and comparable to that of the Type Ia
SN.

Qualitative comparison with the late-time light curves of 
SNe~1990B, 1998bw, and 2002ap indicates a mismatch between the
trend suggested by the last ground-based photometric
points and the HST observation.
SN~1994I shows a sharp downward change of slope in the
exponential tail after $\sim$140 days, not seen in the other
SNe~Ic.
The rate of decay measured from the data between $\sim$80 and 
$\sim$ 140~d after explosion is 43\% slower than that measured 
using the points between $\sim$40 and $\sim$80~d, and the change 
happens very fast, between 80 and 115~d after explosion.
By contrast, the decline slope of SN~1990B is essentially 
constant at these times, that of SN~1998bw changes by 
less than $\sim$28\% between earlier than 80~d after explosion 
and later than 250~d after explosion, and that of SN~2002ap 
changes by $\sim$7\% in this same time period.

If the trend suggested by the last two ground-based photometric 
points were correct, then something fairly dramatic happens to 
the $BV(RI)_C$ light curve between $\sim$140 and $\sim$300~d 
after maximum light.
One possibility is the so-called ``infrared catastrophe'' that 
shifts energy from the optical/near-IR passbands into the 
far-IR \citep{fc89,kandf98}, though it is not expected at 
such relatively early epochs in SNe~Ic.
Another possibility is that the brightness at 300 days after maximum
is affected by a different amount of foreground local extinction, as it would
happen if the ejecta of SN~1994I formed dust.
This is, however, not consistent with the small, or nill, change in broadband
colors.

Another possibility is that the latest ground based photometric points are
affected by contamination from background light from M51.
\citet{retal96} did image matching and subtraction before 
performing photometry, but the background of SN~1994I
was particularly bright and had a steep gradient over the spatial
scales sampled by the PSF of the ground-based images. Thus,
residual contamination in the ground-based images could be present.

\subsection{Comparison with a Simple Model}

When $\gamma$-ray radiative transfer can be approximated by a 
pure absorption process \citep{SandW84}, the radioactive nickel
assumed to be located at the center of the ejecta, accounting
for a small fraction of the total mass (i.e. a point source model
for the $\gamma$-ray emission), and the SN ejecta modeled
by a spherically symmetric 
distribution of matter in free expansion, the optical depth that 
extinguishes the $\gamma$-rays decreases with the inverse of time 
squared \citep[e.g.][]{candw97}.
In this case, the energy deposited by $\gamma$-rays can be 
approximated by
\begin{equation}\label {eq:deposition}
E_\gamma \propto \exp{(-{t/\tau_{\rm Ni}})}   \left\{1 - \exp{[-({T_0/t})^2]}\right\},
\end{equation}
\noindent
where $\tau_{\rm Ni}$ is the e-folding time for the decay 
of $^{56}$\ion{Co}{0}, and $T_0$ is the characteristic time
for a given ejecta structure, which depends on its mass $M$, energy 
$E$, density profile, and the absorption cross section per unit 
mass ($\kappa$) that matter presents to $\gamma$-rays.
For a given density profile, $T_0 \propto (\kappa M^2/E)^{1/2}$.
Soon after maximum light, when radiative equilibrium is established, 
the above assumptions are expected to hold.
Then, in addition, between 100\% and $\sim$96\% of the
radioactive energy is released in $\gamma$-rays, and most of them
are trapped in the ejecta.
In these conditions, $E_\gamma$ gives a good representation of the
energy available to heat the ejecta and
equation \ref{eq:deposition} can be fitted to bolometric light
curves to estimate the parameter $T_0$.
At later times, as the fraction of $\gamma$-rays that escape from the
ejecta increases, the $\sim$4\% of energy released in the form of
positrons needs to be considered.
In what follows, we will consider the two extreme possibilities for
thermalization of the positron annihilation energy: either they are
completely trapped in the ejecta, or they escape.

Assuming that the $BV(RI)_C$ light curve of SN~1994I, between 
$\sim$40 and $\sim$80~d after explosion, is a good representation 
of the bolometric light curve, we fit the light curve to determine
$T_0 \approx 65$~d, if the energy from the $^{56}$\ion{Co}{0} 
positrons is completely deposited, or $T_0 \approx 67$~d, if 
the positrons escape from the ejecta without interacting.
The $BV(RI)_C$ luminosity with these fitted simple models is 
shown in Figure~\ref{fi:models}.

The simple models are very good at fitting the data between 
$\sim$40 and $\sim$80~d after explosion,
suggesting some connection with reality, and they do a 
reasonably good job at $\sim$300~d.
Since the (bolometric) model implies $\sim$50\% more energy than
the $BV(RI)_C$ observation, the data taken at face value would
be consistent either with full trapping of positrons 
from $^{56}$\ion{Co}{0} {\em and} a
small bolometric correction at $\sim$300~d, 
or with a smaller (even null) 
bolometric correction and a more complex process of 
positron transfer that traps $\sim$50\% of their total energy.
It should be noted, though, that the simple model without 
positron trapping falls below the observed light curve.
The simplest interpretation of this observation is that 
SN~1994I did trap a significant fraction of the positrons.

The last two points from ground-based photometry (between $\sim$100
and $\sim$120 days after explosion) are missed by both fits.
We note, in addition, that it is not possible to obtain a good
fit for the simple models of eq.~\ref{eq:deposition} when all
the ground-based observations of the exponential tail (between
$\sim$40 and $\sim$120~d after explosion) are included,
even if the {\em HST} point at 300~d is excluded.
The problem is the fast change in the slope 
of the $BV(RI)_C$ light curve between $\sim$80 
and $\sim$115~d after explosion.
Fitting this break requires including an 
additional free parameter, like
two different values of $T_0$, which
could represent a model of two different mass components
each of them with different expansion properties.
As a comparison, it is possible to successfully fit 
eq.~\ref{eq:deposition} to the exponential decay
of the light curves of all the other SNe~Ic shown in 
Fig.~\ref{fi:BVRI_1}\footnote{We note also that eq. \ref{eq:deposition} 
fails to provide a good fit to the late-time light curve of the 
Type Ia SN~1992A.}.

\subsection{Simple Comparison with Hindsight From the Simple Model} 
\label{sse:norm}

\begin{figure}
\epsscale{0.75}
\plotone{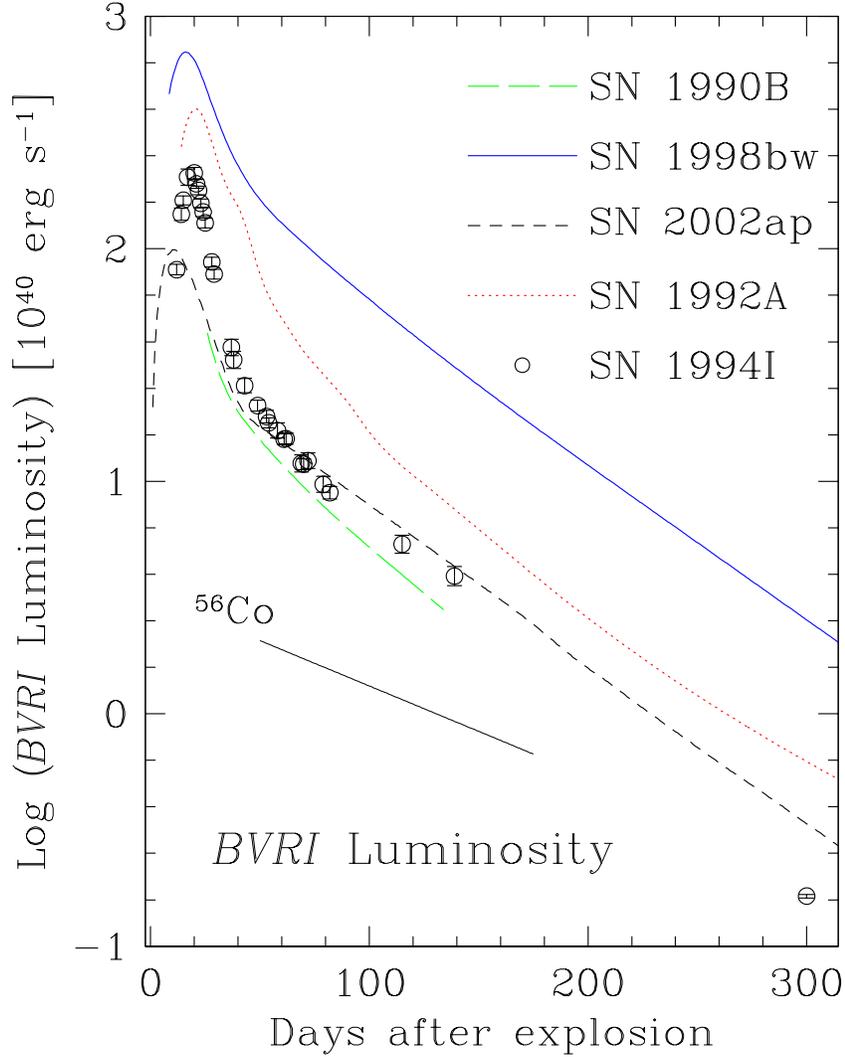}
\caption{ \label{fi:BVRI_1}
$BV(RI)_C$ light curves of the Type~Ic SNe~1994I, 1990B, 
1998bw, and 2002ap, as well as of the Type~Ia SN~1992A.
The light curve of SN~1990B has been displaced downward by
0.5 vertical units for clarity.
Explosion-date estimates are from a theoretical light-curve 
fit for SN~1994I \citep{ietal94}, theoretical spectra and 
light-curve fits for SN~2002ap \citep{metal02}, and the date 
of GRB 980425 for SN~1998bw.
For SN~1990B the explosion date was taken from \citet{cetal01}, 
who estimated it from comparison with SN~1993J, and for 
SN~1992A the explosion date was assumed to be $\sim$20~d before 
maximum light in $B$.
The slope for full trapping of the $^{56}$\ion{Co}{0} decay 
products is also shown.
}
\end{figure}

In \S \ref{sse:simple} we compared the derived
light curves of several SNe.
Although it is illustrative to see how different SN light 
curves behave in a plot like that of Fig.~\ref{fi:BVRI_1},
it is not completely correct to compare the evolution of 
different ejecta on a calendar-time basis.
What we would like to do is to plot the light curves using 
units that are more physically meaningful.

\begin{figure}
\epsscale{0.90}
\plotone{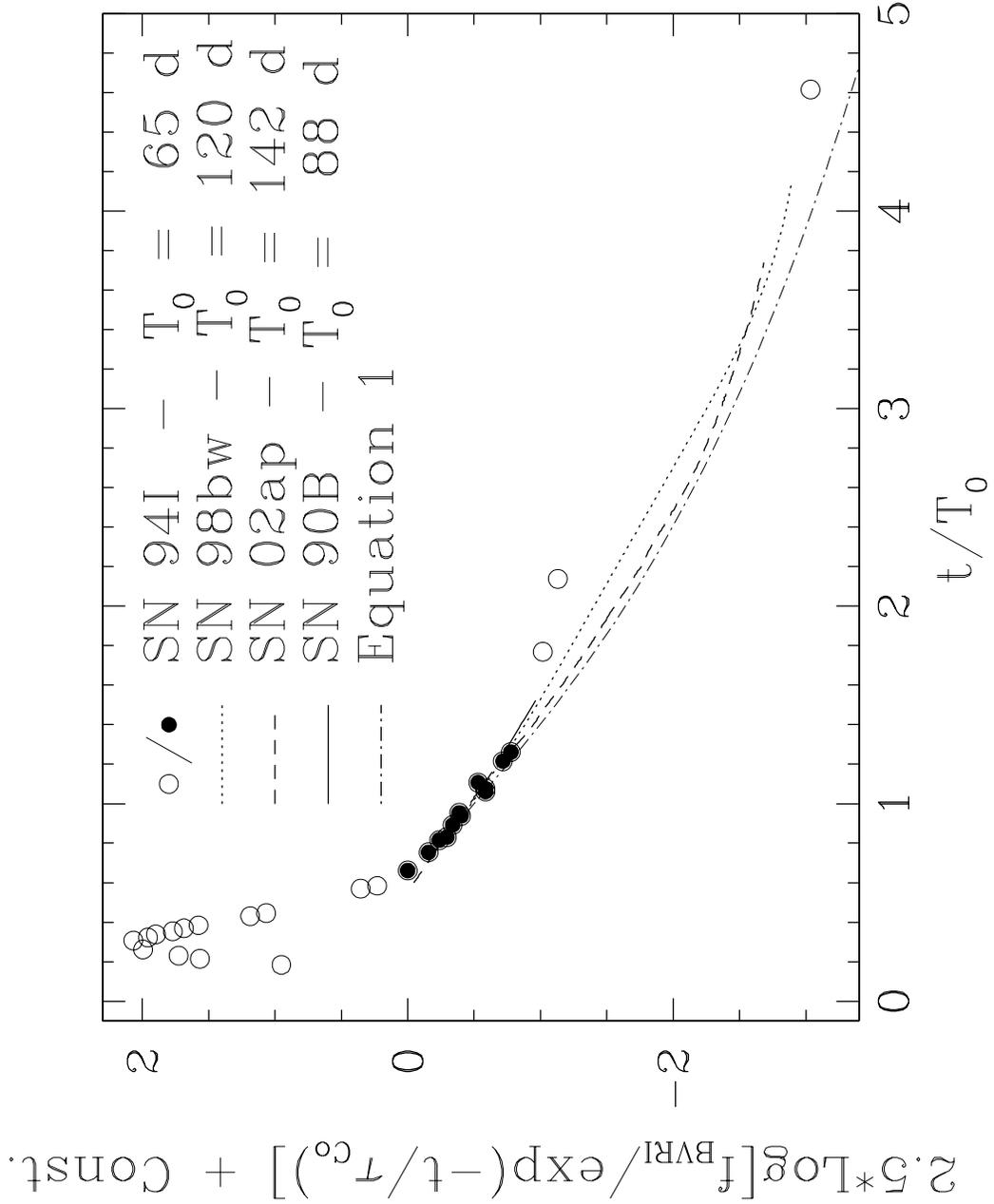}
\caption{ \label{fi:BVRInorm}
$BV(RI)_C$ late-time light curves of SNe~Ic 1994I, 1998bw, 2002ap, 
and 1990B, normalized according to the prescriptions of 
\S~\ref{sse:norm}. Points mark the observations of SN~1994I. 
Solid symbols indicate the region
used to fit the time-scale parameter $T_0$.
A constant was added to the light curves of SNe~1994I, 1998bw, 
2002ap, and 1990B, to make them match the solid points 
of SN~1994I.
}
\end{figure}

From a few tens to a few hundreds of days after explosive 
nucleosynthesis, the evolution of the bolometric light 
curves is governed by two main time scales:
$\tau_{\rm Co}$, the e-folding time for the radioactive 
decay of $^{56}$\ion{Co}{0} into $^{56}$\ion{Fe}{0}, and
$T_0$, the characteristic time for the decrease of the 
$\gamma$-ray optical depth (eq. \ref{eq:deposition}).
They are, respectively, the time scale for energy production 
and the time scale for energy deposition.
This suggests that the natural {\em energy} scale to compare 
late-time light curves of different SNe is
$L_{\rm Bol}(t_n)/\exp{(-{t_n/\tau_{\rm Co}})}$, where $t_n$ 
is measured since nucleosynthesis,
and that the natural scale to compare the time evolution of 
the ejecta is $t_e/T_0$, where $t_e$ should
be measured from the start of free expansion.
Since, at the times of interest, the difference between 
$t_n$ and $t_e$ is negligible, either of them may be used.
We fitted the models of eq. \ref{eq:deposition} to the 
light curves of all SNe~Ic in Fig.~\ref{fi:BVRI_1}, and
then normalized the energy and time axes according to the 
previous prescription.
The result is shown in Figure~\ref{fi:BVRInorm}.

It is reassuring to confirm that such a simple scaling brings 
some ``order'' to Fig.~\ref{fi:BVRI_1}.
When the energy is normalized by the input $^{56}$\ion{Co}{0} 
source, and time by the natural scale for $\gamma$-ray
deposition (responsible for most of the energy available to 
the ejecta from soon after maximum until the supply of 
$^{56}$\ion{Co}{0} is exhausted), all of the late-time light 
curves follow approximately the same track.
It is possible to bring them into close agreement just 
by normalizing the total amount of  $^{56}$\ion{Ni}{0}
ejected by the explosion (i.e. the additive constant
in the logarithm of the energy used in Fig. \ref{fi:BVRI_1}).

This comparison shows that, in units of the natural time scale
of evolution of the $\gamma$-ray optical depth, the late-time point of SN~1994I 
presented in this paper ($\sim 280$~d after maximum) is
even more extreme than those observed for SN~1998bw and SN 2002ap, 
which had light curves extending beyond $\sim$500~d after maximum.
Also, it shows that, in comparison with Eq. \ref{eq:deposition},
the three very late light curves are consistent 
with a flattening after $\sim 3 T_0$, a time at which the 
$\gamma$-ray optical depth becomes negligible, and most of 
the $\gamma$-rays escape into space.
For SN~1994I, the exact time at which the flattening occurs is 
unknown due to the lack of observations.
However, both at $3 T_0 \approx 200$~d, and at 300~d corresponding 
to the last observation, a modest fraction of $^{56}$\ion{Co}{0} 
remains in the ejecta, and the $\sim$4\% of the radiated energy 
that its decay releases in the positron channel becomes a 
significant source.

Regarding the late-time ground-based photometry, the scaling 
of Fig.~\ref{fi:BVRInorm} confirms the arguments given before.
The last two points depart from the common locus, and probably 
overestimate the luminosity of SN~1994I due to background 
contamination.

\subsection{Insights From a Detailed Model}

\citet{ietal94} presented a model of SN 1994I which, without 
fine tuning of the structure, provides a good fit to the early 
part of the $UBV(RI)_C$ light curve, from $\sim 6$~d before up 
to $\sim 40$~d after maximum light.
Their preferred configuration is a C$+$O core that would
result from the evolution of a massive star in an interacting binary
system that underwent two mass-transfer episodes.
Depending on the initial parameters of the binary, at the moment of 
explosion the pair consists of the presupernova and a neutron 
star, white dwarf, or low-mass main-sequence companion 
\citep{netal94}.

\citet{ietal94} fitted the light curves of SN~1994I provided 
by \citet{schmidt94} with the explosion of a star that entered 
the main sequence with $\sim 15 M_\sun$, became a 
$\sim 4.0 M_\sun$ He star after the first mass transfer,
and a 2.1 $M_\sun$ C$+$O core after the second (model CO21).
A simultaneous fit of the monochromatic and $UBV(RI)_C$ curves
required an interstellar reddening of $E(B-V) = 0.45$ mag, and 
a distance of 6.92 Mpc ($m-M = 29.2$ mag), in good agreement 
with the values used to determine the $BV(RI)_C$ light curve 
presented here.

Would model CO21 fit the very late-time {\em HST} photometry?
The question is relevant because previous attempts at fitting 
SN light curves over long time spans have uncovered
a problem. When the models are good at fitting the maximum 
light they tend to be too dim on the tail; conversely,
if they are good at fitting the late-time decay, they tend 
to give a maximum broader and/or brighter than observed.
Solving this problem has typically meant introducing free 
parameters in the density structure of the model
\citep{eandw88,tetal06}.

Based on the explosion model CO21 \citep{ietal94},
and using the same radiative transfer code,
we calculated the monochromatic light curves, but
here we extend the computations to 300 days after
explosion.  
The assumptions and approximations used in the code
are described in more detail in \citep{ietal94}
and  \citep{hetal93}.
A hypothesis relevant to our discussion below,
is that all positrons are thermalized locally
and cannot escape.
Because of the limitations of the code 
in the treatment of forbidden lines and their time dependence, 
we assumed the energy distribution to be unchanged after day 
120.
This assumption is consistent with the observations 
because the color indices scarcely change after about 100~d 
(Table~\ref{ta:groundphot} and Fig. \ref{fi:colors}).
The assumption prevents us from testing overall migration
of flux toward the IR, as may result from the formation of dust.
The evolution of the colors, however, suggest that neither
dust formation, nor other change in the foreground extinction
by matter in the circumstellar region, do occur in this SN.

From the monochromatic fluxes, we calculated the magnitudes
in each relevant $HST$ passband.
We checked that the synthetic $F439W$ and $F555W$ values agree
with the $B$ and $V$ of \citet{ietal94} within a few hundredths
of a magnitude.
The small differences can be attributed to adjustments in 
the frequency grid and differences in the {\em HST} filter 
functions and the Bessell system, since $S$-corrections were
not applied in this test.

\begin{figure}
\epsscale{0.80}
\plotone{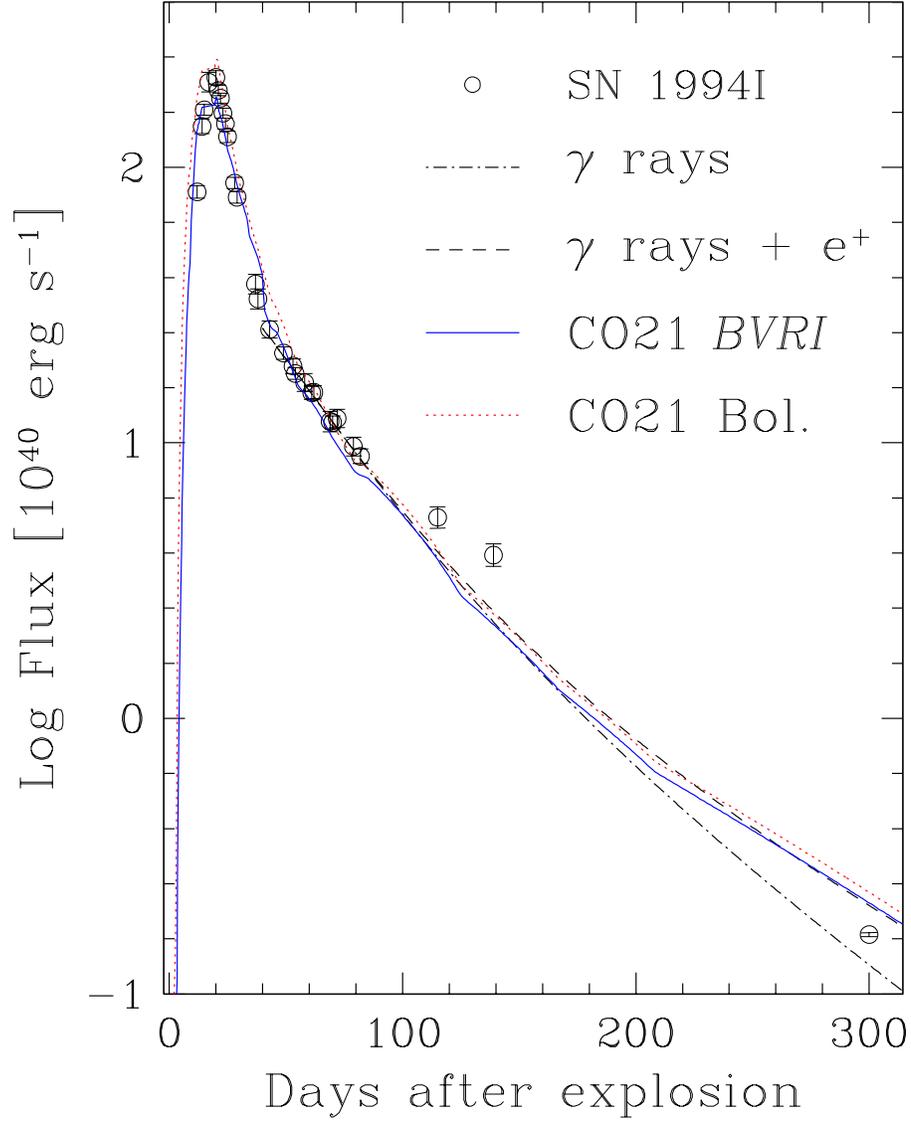}
\caption{ \label{fi:models}
$BV(RI)_C$ light curve of SN~1994I, the simple models given by 
eq.~\ref{eq:deposition}, and model CO21.
}
\end{figure}

From the monochromatic fluxes at the effective wavelength of
the passbands, we built a $BV(RI)_C$ light curve using the same
procedure as in the observations (see \S \ref{se:lcurves}).
We compare the theoretical light curves of model CO21 with 
the observations in Fig.~\ref{fi:models}.
Both the bolometric and $BV(RI)_C$ luminosities are plotted.
They are very similar, indicating that the model concentrates 
most of the energy within the passbands observed.
Overall, the model luminosities agree well with the observations, 
especially in the post-maximum decay, although some details
merit further comment.

At maximum, the model results dimmer than the observations, while
the same model fits better the $UBV(RI)_C$ in \citet{ietal94}.
One reason for that is the distance.
The observed $BV(RI)_C$ light curve presented here is based on an
observed distance of 8.39 Mpc \citep{fcj97}, while, for the same
foreground extinction,
CO21 is consistent with a distance 1.47 Mpc shorter in \citet{ietal94}.
The shorter distance will help our model $BV(RI)_C$ light curve to fit
the maximum better, but will give a brighter than required post maximum
decay.
This highlights the importance of using extensive datasets when comparing
models to observations.

The $BV(RI)_C$ light curve of CO21 shown in Fig.~\ref{fi:models}
does fit most of the observed $BV(RI)_C$ light curve of SN~1994I.
First, the distance independent width of the observed maximum, and then
the distance dependent brightness of the post-maximum decline and exponential decay.
Fitting the peak-to-tail contrast will require some additional fine tunning.

A possible hint is that the model $UBV(RI)_C$ light curve does a better job
at fitting the peak-to-tail contrast of the $BV(RI)_C$ observations, indicating
that the model spectrum overestimates the fraction of total flux emerging in the
ultraviolet at early times.
In any case, the problem of CO21 is opposite to the traditional peak-to-tail
contrast problem: if the maximum of the $BV(RI)_C$ model is arbitrarily shifted to match
the observed one, then the model light curve results brighter than the observed,
not dimmer.
Studying the peak-to-tail contrast problem in SN light curves is beyond the scope of this
paper, so we will let the issue rest there.

The late-time tail of the detailed model, which includes a 
realistic approximation to the $\gamma$-ray transfer and
deposition, follows closely the simple deposition function 
given by eq. \ref{eq:deposition}, with full trapping of the 
positrons from the decay of $^{56}$\ion{Co}{0}.
Since CO21 also assumes full trapping of positrons, this is 
an indication that the $\gamma$-ray transfer is well represented
by the simple approximations that went into 
eq. \ref{eq:deposition}.
Hence, the conclusion stated above that about half of the 
positrons are able to deposit their energy within the ejecta 
is supported by the more elaborate model as well.

\section{Conclusions} \label{se:conclusions}

Through the {\em HST} GO SINS program, we observed SN~1994 
with WFPC2 when the SN was about 280~d past maximum light.
We did PSF photometry on the resulting images using 
HSTPhot \citep{dol00}.

We collected published ground-based photometry of SN~1994I, 
applied a reddening correction, and transformed the results 
to the space-based {HST} WFPC2 photometric system.
Both the reddening correction and the photometric transformation 
were based on the extensive spectroscopic database available
for the SN \citep{fetal95, cetal96b}.
The resulting light curves in the $F439W$, $F555W$, $F675W$, 
and $F814W$ passbands span from $\sim$7~d before to $\sim$280~d
after maximum light.

Using the zero points and effective wavelengths of \citet{hol95}, we
transformed the multicolor photometry to monochromatic flux, and 
then integrated the flux over the available wavelength range to 
construct a ``$BV(RI)_C$'' light curve.
We compared the $BV(RI)_C$ light curve of SN~1994I with those of
other SNe~Ic, and, based on a simplified model of the ejecta, 
proposed a scaling in energy and time that provides a more 
insightful comparison of the exponential tails of different SNe.
The light curves that reach times larger than $\sim 3 T_0$, 
where $T_0$ is the characteristic time for the
decay of the $\gamma$-ray optical depth, show a clear flattening.
For rapidly evolving SNe, like SN~1994I, a simple interpretation 
of this flattening is that the small percentage of energy that the
decay of $^{56}$\ion{Co}{0} releases in the form of positrons 
becomes a relevant thermal energy source for the ejecta.

Model CO21, originally presented by \citet{ietal94}, does a 
reasonable job at fitting both the $BV(RI)_C$ and monochromatic 
light curves at early times.
We extended the original light-curve calculations up to day 300. 
We note that the absolute $BV(RI)_C$ luminosity agrees well 
with the observations of SN 1994I up to about 80~d after maximum 
light, but overestimates the flux by $\sim 50$\% at day 300.

Models with full positron trapping indicate that some energy 
was missed by the observations, but models without positron 
trapping indicate an observed excess of energy.
The missing (or excess) energy amounts to $\sim$50\% of that 
produced by the annihilation of $^{56}$\ion{Co}{0} positrons.
Although models with no positron trapping {\em and} a migration  
of energy toward the mid-to-far IR (e.g., by dust formation) 
cannot be ruled out, a simpler intepretation of these late-time 
observations is that the ejecta of SN~1994I trapped $\sim$50\% 
of the positrons roughly 300~d after the explosion.

Our interpretation of the $BV(RI)_C$ light curve
is consistent with the SN evolving as a freely expanding
envelope whose optical depth to $\gamma$-rays decreases,
very approximately, with the inverse of time squared.
Simple models that treat the deposition of $\gamma$-rays 
as a pure absorption process, and the more detailed model 
CO21, agree on this.
SN~1994I does not need a model with two dynamically distinct 
components, like the one that \citet{tetal06} propose for 
SN~2002ap.

We conclude that our very late-time $BV(RI)_C$ photometry
represents most of the energy that the SN produced at
this time, implying that the late-time bolometric correction of 
this rapidly evolving SN~Ic was not large.
This, in turn, would mean that the ``infrared catastrophe'' in 
this SN does not happen until phases later than $\sim$300~d.

Although model CO21 was not finely tuned to closely match 
the observed light curves of SN~1994I, it helps to show that
the late-time photometry presented here is extreme in terms 
of the $\gamma$-ray optical depth, and agrees with the simple
models of eq. \ref{eq:deposition} in that the dominant source 
of late-time energy originates in positrons.
A model built specifically to represent this SN could be 
used to establish a meaningful constraint on the fraction 
of positrons actually trapped in the ejecta, and whether
(and when) there is a migration of photons from the optical 
and near-IR to the IR region of the spectrum.

\acknowledgments A.C. thanks the Department of Astronomy of 
the University of Texas at Austin for support from a W. M. Cox 
Visitor Grant and from the Samuel T. and Fern Yanagisawa 
Regents Professorship in Astronomy, and acknowledges the  
support of CONICYT, Chile, under grants FONDECYT 1051061 and
FONDAP 15010003. 
The {\it HST} SINS program was supported by 
NASA through grants GO--2563 and GO--5777 from the Space Telescope Science
Institute, which is operated by the Association of Universities 
for Research in Astronomy (AURA), Inc., under NASA contract 
NAS 5--26555.
J.C.W. was supported in part by NASA Grant NNG04GL00G and NSF 
Grant AST--0406740. A.V.F. is grateful for the support of
NSF grant AST--0607485.
This research made use of the SUSPECT online database 
of SN spectra (http://bruford.nhn.ou.edu/$\sim$suspect/).

\clearpage

\clearpage

\clearpage

\clearpage

\clearpage

\clearpage

\clearpage

\clearpage

\begin{deluxetable}{llrcrr}
\tablewidth{0pt}
\tablecaption{Late-Time {\em HST} Photometry of SN~1994I\label{ta:phot}}
\tablehead{
\colhead{{\em HST} Dataset} &
\colhead{JD\tablenotemark{a}} &
\colhead{Exp.\tablenotemark{b}} &
\colhead{Passband} &
\colhead{{\em HST} Mag.} &
\colhead{Extinction\tablenotemark{c}} 
}
\startdata
U2LQ0201T/2T & 733.21 & 1400 & $F439W$ & 21.63 $\pm$ 0.02 & 1.89 $\pm$ 0.033   \hfill \\
U2LQ0203T & 733.22 &  600 & $F555W$ & 21.03 $\pm$ 0.01 & 1.45 $\pm$ 0.030  \hfill \\
U2LQ0204T & 733.25 &  600 & $F675W$ & 20.21 $\pm$ 0.01 & 1.12 $\pm$ 0.010  \hfill \\
U2LQ0205T & 733.26 &  600 & $F814W$ & 19.69 $\pm$ 0.01 & 0.87 $\pm$ 0.010  \hfill \\
\enddata
\tablenotetext{a}{Mean Julian Date of total exposure minus 2,449,000. For 
phase, subtract an additional 433.00.}
\tablenotetext{b}{Total exposure time in the passband (in seconds). Two 
$F439W$ exposures of 700~s each were combined in one image.}
\tablenotetext{c}{Interstellar extinction in the host galaxy, in the 
{\em HST} passband.}
\end{deluxetable}

\begin{deluxetable}{lllllllllrrrr}
\rotate
\tablewidth{0pt}
\tablecaption{Ground-Based Photometry of SN~1994I\label{ta:groundphot}}
\tablehead{
\colhead{JD\tablenotemark{a}} &
\colhead{$B$\tablenotemark{b}} &
\colhead{$V$\tablenotemark{b}} &
\colhead{$R_C$\tablenotemark{b}} &
\colhead{$I_C$\tablenotemark{b}} &
\colhead{$A_B$\tablenotemark{c}} &
\colhead{$A_V$\tablenotemark{c}} &
\colhead{$A_{R_C}$\tablenotemark{c}} &
\colhead{$A_{I_C}$\tablenotemark{c}} &
\colhead{$SC_B$\tablenotemark{d}} &
\colhead{$SC_V$\tablenotemark{d}} &
\colhead{$SC_{R_C}$\tablenotemark{d}} &
\colhead{$SC_{I_C}$\tablenotemark{d}}
}
\startdata
445 & 14.41 $\pm$ 0.03 & 14.04 $\pm$ 0.02 & 13.87 $\pm$ 0.02 & 13.77 $\pm$ 0.01 & 1.83 & 1.40 & 1.17 & 0.86 & -0.03 & -0.04 & 0.05 & -0.01 \\
447 & 13.97 $\pm$ 0.08 & 13.39 $\pm$ 0.02 & 13.21 $\pm$ 0.02 & 13.09 $\pm$ 0.02 & 1.83 & 1.40 & 1.17 & 0.86 & -0.04 & -0.04 & 0.06 & -0.01 \\
448 & 13.95 $\pm$ 0.02 & 13.20 $\pm$ 0.01 & 13.02 $\pm$ 0.01 & 12.88 $\pm$ 0.01 & 1.83 & 1.40 & 1.17 & 0.86 & -0.05 & -0.04 & 0.06 & -0.01 \\
450 & 13.68 $\pm$ 0.21 & 12.99 $\pm$ 0.02 & 12.72 $\pm$ 0.02 & 12.60 $\pm$ 0.02 & 1.83 & 1.40 & 1.17 & 0.86 & -0.06 & -0.04 & 0.06 & 0.00 \\
453 & 13.81 $\pm$ 0.07 & 12.91 $\pm$ 0.02 & 12.62 $\pm$ 0.02 & 12.49 $\pm$ 0.02 & 1.83 & 1.40 & 1.17 & 0.86 & -0.07 & -0.04 & 0.06 & 0.00 \\
454 & 14.05 $\pm$ 0.02 & 13.02 $\pm$ 0.01 & 12.72 $\pm$ 0.01 & 12.52 $\pm$ 0.01 & 1.83 & 1.40 & 1.17 & 0.86 & -0.07 & -0.04 & 0.06 & 0.00 \\
455 & 14.15 $\pm$ 0.02 & 13.11 $\pm$ 0.01 & 12.76 $\pm$ 0.01 & 12.53 $\pm$ 0.01 & 1.83 & 1.40 & 1.17 & 0.86 & -0.08 & -0.04 & 0.06 & 0.00 \\
456 & 14.44 $\pm$ 0.01 & 13.23 $\pm$ 0.01 & 12.86 $\pm$ 0.01 & 12.61 $\pm$ 0.01 & 1.83 & 1.40 & 1.17 & 0.86 & -0.08 & -0.04 & 0.06 & 0.00 \\
457 & 14.56 $\pm$ 0.02 & 13.34 $\pm$ 0.01 & 12.94 $\pm$ 0.02 & 12.64 $\pm$ 0.01 & 1.83 & 1.40 & 1.17 & 0.86 & -0.08 & -0.04 & 0.06 & 0.00 \\
458 & 14.75 $\pm$ 0.02 & 13.46 $\pm$ 0.02 & 13.04 $\pm$ 0.02 & 12.72 $\pm$ 0.02 & 1.83 & 1.41 & 1.17 & 0.86 & -0.08 & -0.04 & 0.06 & 0.00 \\
461 & 15.28 $\pm$ 0.03 & 13.93 $\pm$ 0.02 & 13.41 $\pm$ 0.02 & 13.00 $\pm$ 0.02 & 1.83 & 1.41 & 1.17 & 0.86 & -0.09 & -0.04 & 0.06 & 0.00 \\
462 & 15.40 $\pm$ 0.04 & 14.08 $\pm$ 0.02 & 13.53 $\pm$ 0.03 & 13.10 $\pm$ 0.02 & 1.83 & 1.41 & 1.17 & 0.85 & -0.09 & -0.03 & 0.06 & 0.00 \\
470 & 16.00 $\pm$ 0.32 & 14.92 $\pm$ 0.02 & 14.37 $\pm$ 0.03 & 13.79 $\pm$ 0.01 & 1.84 & 1.41 & 1.17 & 0.85 & -0.09 & -0.03 & 0.07 & 0.01 \\
471 & 15.97 $\pm$ 0.30 & 15.11 $\pm$ 0.02 & 14.50 $\pm$ 0.02 & 13.96 $\pm$ 0.02 & 1.84 & 1.41 & 1.17 & 0.85 & -0.09 & -0.03 & 0.07 & 0.01 \\
476 & 16.28 $\pm$ 0.10 & 15.41 $\pm$ 0.05 & 14.80 $\pm$ 0.07 & 14.12 $\pm$ 0.01 & 1.84 & 1.41 & 1.17 & 0.85 & -0.08 & -0.02 & 0.07 & 0.01 \\
482 & 16.61 $\pm$ 0.08 & 15.56 $\pm$ 0.02 & 15.03 $\pm$ 0.03 & 14.37 $\pm$ 0.02 & 1.84 & 1.41 & 1.17 & 0.85 & -0.06 & -0.02 & 0.07 & 0.01 \\
486 & 16.56 $\pm$ 0.11 & 15.71 $\pm$ 0.03 & 15.20 $\pm$ 0.04 & 14.53 $\pm$ 0.01 & 1.84 & 1.41 & 1.17 & 0.85 & -0.05 & -0.01 & 0.08 & 0.01 \\
487 & 16.78 $\pm$ 0.04 & 15.76 $\pm$ 0.02 & 15.24 $\pm$ 0.03 & 14.54 $\pm$ 0.02 & 1.84 & 1.41 & 1.17 & 0.85 & -0.05 & -0.01 & 0.08 & 0.01 \\
491 & 16.58 $\pm$ 0.16 & 15.90 $\pm$ 0.03 & 15.41 $\pm$ 0.07 & 14.62 $\pm$ 0.02 & 1.84 & 1.41 & 1.17 & 0.85 & -0.04 & -0.01 & 0.08 & 0.01 \\
494 & 16.88 $\pm$ 0.07 & 15.94 $\pm$ 0.02 & 15.45 $\pm$ 0.03 & 14.75 $\pm$ 0.02 & 1.84 & 1.41 & 1.17 & 0.85 & -0.03 & -0.01 & 0.08 & 0.02 \\
495 & 16.86 $\pm$ 0.09 & 15.89 $\pm$ 0.02 & 15.49 $\pm$ 0.02 & 14.80 $\pm$ 0.03 & 1.84 & 1.41 & 1.17 & 0.85 & -0.03 & -0.01 & 0.08 & 0.02 \\
502 & 17.21 $\pm$ 0.14 & 16.18 $\pm$ 0.07 & 15.73 $\pm$ 0.09 & 15.00 $\pm$ 0.03 & 1.84 & 1.41 & 1.17 & 0.85 & -0.02 & 0.00 & 0.08 & 0.02 \\
503 & 17.06 $\pm$ 0.06 & 16.24 $\pm$ 0.02 & 15.75 $\pm$ 0.04 & 15.04 $\pm$ 0.02 & 1.84 & 1.41 & 1.17 & 0.85 & -0.02 & 0.00 & 0.08 & 0.02 \\
505 & 16.93 $\pm$ 0.17 & 16.21 $\pm$ 0.04 & 15.74 $\pm$ 0.06 & 15.08 $\pm$ 0.02 & 1.84 & 1.41 & 1.17 & 0.85 & -0.01 & 0.00 & 0.09 & 0.02 \\
512 & 17.21 $\pm$ 0.18 & 16.48 $\pm$ 0.04 & 15.99 $\pm$ 0.06 & 15.30 $\pm$ 0.02 & 1.84 & 1.42 & 1.17 & 0.85 & -0.01 & 0.01 & 0.09 & 0.02 \\
515 & 17.48 $\pm$ 0.08 & 16.58 $\pm$ 0.04 & 15.97 $\pm$ 0.03 & 15.37 $\pm$ 0.05 & 1.84 & 1.42 & 1.17 & 0.85 & 0.00 & 0.01 & 0.09 & 0.02 \\
548 & 18.05 $\pm$ 0.13 & 17.23 $\pm$ 0.08 & 16.46 $\pm$ 0.08 & 15.95 $\pm$ 0.05 & 1.85 & 1.42 & 1.17 & 0.86 & 0.00 & 0.03 & 0.11 & 0.03 \\
572 & 18.10 $\pm$ 0.18 & 17.60 $\pm$ 0.06 & 16.93 $\pm$ 0.09 & 16.34 $\pm$ 0.05 & 1.85 & 1.42 & 1.17 & 0.86 & 0.00 & 0.04 & 0.12 & 0.04 \\
\enddata
\tablenotetext{a}{Mean Julian Date minus 2,449,000. For phase, subtract an 
additional 433.00.}
\tablenotetext{b}{Photometry from Tables 4, 5, or 6 of 
{\protect{\citet{retal96}}}.}
\tablenotetext{c}{Interstellar extinction in the host galaxy, in the 
given ground-based passband.
Mean uncertainties (mag) are
$\delta A_B    = 0.060$,
$\delta A_V    = 0.010$,
$\delta A_{Rc} = 0.003$, and
$\delta A_{Ic} = 0.003$.
}
\tablenotetext{d}{$S$--correction to be subtracted from the 
ground-based magnitude to transform to the space-borne 
equivalent passband (see Table \protect{\ref{ta:phot}}).
Mean uncertainties (mag) are
$\delta SC_B    = 0.033$,
$\delta SC_B    = 0.041$,
$\delta SC_{Rc} = 0.053$, and
$\delta SC_{Ic} = 0.014$.
}
\end{deluxetable}

\end{document}